\documentclass[conference,anonymous]{IEEEtran}
\IEEEoverridecommandlockouts

\usepackage[utf8]{inputenc}
\usepackage{cite}
\usepackage{amsmath,amssymb,amsfonts}
\usepackage{graphicx}
\usepackage{textcomp}
\usepackage{xcolor}
\usepackage{hyperref}
\usepackage{url}
\usepackage{booktabs}
\usepackage{multirow}
\usepackage{balance}

\def\BibTeX{{\rm B\kern-.05em{\sc i\kern-.025em b}\kern-.08em
    T\kern-.1667em\lower.7ex\hbox{E}\kern-.125emX}}

\begin{document}

\title{Toward Affordable and Non-Invasive Detection of Hypoglycemia: A Machine Learning Approach}

\author{\IEEEauthorblockN{Lawrence Obiuwevwi, Krzysztof J. Rechowicz, Vikas Ashok, Sampath Jayarathna}
\IEEEauthorblockA{\textit{Department of Computer Science, Virginia Digital Maritime Center (VDMC)}\\
\textit{Old Dominion University}, Norfolk, VA, USA \\
\{lobiu001@odu.edu, krechowi@odu.edu, vganjigu@cs.odu.edu, sampath@cs.odu.edu\}}
}

\maketitle

\begin{abstract}
Diabetes mellitus is a growing global health issue, with Type 1 Diabetes (T1D) requiring constant monitoring to avoid hypoglycemia. Although Continuous Glucose Monitors (CGMs) are effective, their cost and invasiveness limit access, particularly in low-resource settings. This paper proposes a non-invasive method to classify glycemic states using Galvanic Skin Response (GSR), a biosignal commonly captured by wearable sensors. We use the merged OhioT1DM 2018 and 2020 datasets to build a machine learning pipeline that detects hypoglycemia (glucose less than 70 mg/dL) and normoglycemia (greater than 70 mg/dL) with GSR alone. Seven models are trained and evaluated: Random Forest, XGBoost, MLP, CNN, LSTM, Logistic Regression, and K-Nearest Neighbors. Validation sets and 95\% confidence intervals are reported to increase reliability and assess robustness. Results show that the LSTM model achieves a perfect hypoglycemia recall (1.00) with an F1-score confidence interval of [0.611–0.745], while XGBoost offers strong performance with a recall of 0.54 even under class imbalance. This approach highlights the potential for affordable, wearable-compatible glucose monitoring tools suitable for settings with limited CGM availability using GSR data.
\end{abstract}

\begin{IEEEkeywords}
Hypoglycemia Detection, Galvanic Skin Response, Non-Invasive Monitoring, Wearables, Machine Learning, Confidence Intervals.
\end{IEEEkeywords}

\section{Introduction}

Diabetes affects over 537 million people globally, with projections rising to 783 million by 2045, particularly in low and middle income countries \cite{idf2021, di_filippo_non-invasive_2023}. For those with T1D, constant glucose level tracking is vital to prevent complications from hypoglycemia and hyperglycemia \cite{mathew_hypoglycemia_2025}. While CGMs provide real-time data, their high cost and maintenance limit use in underserved areas \cite{lehmann_noninvasive_2023, di_filippo_non-invasive_2023}.

Wearable biosensors offer a promising alternative. Galvanic Skin Response (GSR), or electrodermal activity (EDA), reflects changes in skin conductance modulated by the sympathetic nervous system \cite{meijer_electrodermal_2023, znidaric_electrodermal_2023}. GSR has shown potential in stress and fatigue detection, and recent studies suggest a physiological link between hypoglycemia and sympathetic arousal \cite{polak_analysis_2024, hongn_wearable_2025}. Unlike invasive sensors, GSR can be collected passively via smartwatches, making it attractive for large-scale, low-cost monitoring \cite{ribeiro_novel_2024}. Though GSR has been explored in emotion and stress contexts \cite{yu_exploring_2024}, its application in glycemic monitoring remains limited.

This study investigates GSR as a stand-alone signal for classifying glucose states. We use the merged OhioT1DM 2018 and 2020 datasets \cite{marling_ohiot1dm_2020}, which contain synchronized GSR and glucose data. The decision to focus solely on GSR and glucose was driven by our goal of enabling low-cost deployment via commercial wearables, most of which currently only support GSR and lack multi-sensor integration. Other physiological features in the dataset (e.g., Heart rate, temp) were excluded to isolate GSR’s standalone potential and reduce model complexity for real-world applications.

Our pipeline aligns these signals and evaluates seven models—Random Forest (RF), XGBoost, Multilayer Perceptron (MLP), Convolutional Neural Network (CNN), Long Short-Term Memory (LSTM), Logistic Regression, and K-Nearest Neighbors (KNN). These models were chosen to balance interpretability, computational cost, and the ability to learn temporal and nonlinear patterns. We include both classical and deep learning methods to compare performance across paradigms.

Given the rarity of hypoglycemic events (4.1\% of samples), we address class imbalance using weighted loss functions, and stratified train/validation/test splits (80/10/10) to ensure fair representation and reliable evaluation. Confidence intervals (95\%) are reported for accuracy, recall, and F1-score on the test set to assess statistical robustness. The 80/10/10 split was chosen to provide sufficient training data while enabling independent validation and testing, an essential step for generalization given the imbalanced nature of the task.

Our main contributions are:
\begin{itemize}
    \item A GSR-only machine learning framework for binary glucose state classification.
    \item Evaluation of seven models using a merged, real-world T1D dataset with synchronized biosignals.
    \item Emphasis on hypoglycemia recall, class imbalance mitigation, and performance confidence intervals.
    \item Demonstration that LSTM (recall = 1.00, F1 = 0.68, CI$_{\text{F1}}$: 0.611–0.745) and XGBoost (recall = 0.54) models show promise for deployment in wearable-compatible glucose alert systems.
\end{itemize}

The findings show that GSR, when processed effectively, holds promise for non-invasive glucose monitoring, offering a viable, accessible tool to support diabetes care in diverse settings.

\section{Related Work}

The application of machine learning (ML) to diabetes prediction, diagnosis, and management has evolved significantly in the past decade. A broad range of techniques, from traditional ensemble methods to advanced deep learning architectures, have been employed to forecast blood glucose levels, identify glycemic patterns, and anticipate hypoglycemic events \cite{mujahid_machine_2021}. These efforts are increasingly focused on enabling non-invasive, real-time, and personalized monitoring for both clinical and home-based care.

\subsection{Data-Driven Glucose Forecasting}
Machine learning has shown strong potential in forecasting blood glucose levels. Woldaregay et al. emphasized personalization and temporal modeling as critical factors for accuracy in T1D glucose prediction \cite{woldaregay_data-driven_2019}. Nemat et al. found deep learning models, especially recurrent networks, outperform traditional methods given rich sequential data \cite{nemat_data-driven_2024}. Neumann et al. integrated physical activity and meals into adaptive, real-world models \cite{neumann_data-driven_2025}. Neamtu et al. demonstrated that small, individualized datasets can still yield reliable predictions in pediatric cases \cite{neamtu_predicting_2023}. Additionally, Butt et al. showed that applying feature transformation improves both performance and interpretability \cite{butt_feature_2023}.

\subsection{Hypoglycemia Event Detection and Classification}

Hypoglycemia is one of the most dangerous and immediate complications in diabetes care. Fleischer et al. implemented an ensemble model using CGM data to provide early warnings for impending hypoglycemia \cite{fleischer_hypoglycemia_2022}. Jayarathna et al. demonstrated the potential of eye movement metrics, specifically saccade velocity patterns, as indicators of nocturnal hypoglycemia by linking reduced saccadic variability to low blood glucose events detected via CGM \cite{jayarathna2025nocturnal}. Faccioli et al. proposed a prediction-funnel alarm system for timely intervention in online glucose forecasting \cite{faccioli_combined_2023}. Tsichlaki et al. documented the shift from rule-based systems to probabilistic and sequence-aware models that learn individual responses to physiological stimuli \cite{tsichlaki_type_2022}.

\subsection{Electrodermal Activity in Health Monitoring}

EDA (or GSR) reflects autonomic nervous system activity and has traditionally been used for psychological arousal and stress monitoring. Polak et al. explored the correlation between GSR and blood glucose levels in diabetic patients, finding consistent skin conductance patterns preceding hypoglycemia events \cite{polak_analysis_2024}. Meijer et al. showed that EDA could serve as a continuous monitor for physiological instability \cite{meijer_electrodermal_2023}. Znidaric et al. proposed a multimodal framework combining GSR and HRV for early detection of Type 2 diabetes complications \cite{znidaric_electrodermal_2023}. Hongn et al. revealed GSR changes under exercise and anxiety, supporting its physiological relevance beyond emotion tracking \cite{hongn_wearable_2025}. Yu et al. demonstrated GSR’s sensitivity to systemic stress responses \cite{yu_exploring_2024}.

\subsection{Recent Advances in Wearable Hypoglycemia Detection}

Recent studies support the use of GSR for non-invasive hypoglycemia detection. Thong et al. trained machine learning models on smartwatch-collected heart rate, HRV, and tonic EDA, achieving an AUROC of 0.76 and + or - 0.07 for hypoglycemia detection using only wearable data \cite{thong2023noninvasive}. Maritsch et al. demonstrated that GSR and heart rate signals from smartwatches could detect hypoglycemia under real-world stress conditions like driving \cite{maritsch2023smartwatches}. Rubega et al. showed that EEG complexity could distinguish hypoglycemia events with approximately 90\% accuracy, reinforcing the power of physiological signals for glucose monitoring tasks \cite{rubega2020eeg}.

\subsection{Electrodermal Activity in Diabetes and Hypoglycemia Prediction}
Only a few studies have attempted to directly integrate GSR into diabetes management pipelines. Mendez et al. used smartwatch-based GSR and accelerometry data for detecting nocturnal hypoglycemia, showing the promise of passive biosignals for in-home monitoring \cite{mendez_toward_2025}. Lehmann et al. also demonstrated that ML models trained on GSR and motion data could reliably classify hypoglycemia events without requiring CGM input \cite{lehmann_noninvasive_2023}.

Despite these advances, challenges persist in isolating hypoglycemia-induced GSR signals from confounders like temperature, stress, and individual variability. Our study addresses these gaps through rigorous evaluation of GSR-only models on a merged, time-aligned dataset, reporting confidence intervals and focusing on recall to support real-world reliability.

\section{Materials and Methods}

In this section we detailed the dataset used, preprocessing pipeline, labeling strategy, and the machine learning models applied to build the glucose state classification framework using GSR alone.

\begin{figure}[ht]
\centering
    \includegraphics[width=\linewidth]{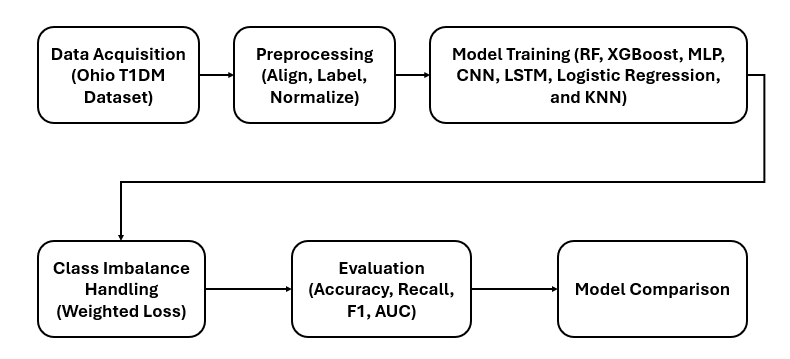}
    \caption{
    End-to-End Machine Learning Workflow for Glucose State Prediction:
    This flowchart outlines the complete process from data acquisition (OhioT1DM) to model comparison, highlighting preprocessing, training, imbalance handling, and performance evaluation steps.}
    \label{fig:GSR and Glucose Signals for Model Training and Testing}
\end{figure}

\subsection{Dataset Description}

We used the OhioT1DM datasets, publicly released in 2018 and updated in 2020, for this study \cite{marling_ohiot1dm_2020}. These datasets comprise multi-weeks physiological, behavioral, and metabolic data from 12 individuals diagnosed with Type 1 Diabetes (T1D). Participants were monitored continuously using a combination of Continuous Glucose Monitors, wearable biosensors (Microsoft Band 2), and self-reported event logs for meals, insulin dosage, and exercise routines.

The dataset includes:
\begin{itemize}
    \item Interstitial glucose measurements recorded every 5 minutes by CGMs.
    \item GSR captured at regular intervals by wrist-worn sensors.
    \item Additional signals such as heart rate, accelerometry, and skin temperature.
\end{itemize}

For this study, only GSR and glucose values were retained. This decision was motivated by our aim to evaluate GSR as a standalone non-invasive biomarker for hypoglycemia detection, simulating realistic wearable sensor limitations. The 2018 and 2020 versions were merged, yielding a total of 401,471 synchronized time steps. This long-term, continuous dataset offers one of the most comprehensive resources for evaluating physiological markers of glycemic changes in free-living conditions.

\subsection{Data Preprocessing}

Due to differing sampling rates between CGM and GSR signals, data alignment was a critical preprocessing step. GSR values were interpolated to match the 5-minute CGM interval, and mean aggregation was applied where GSR had higher frequency.

Outliers were removed using the interquartile range (IQR) method. A low-pass Butterworth filter was used to smooth noisy conductance values \cite{ribeiro_novel_2024}. To correct for baseline variability across subjects, GSR signals were standardized per participant using z-score normalization.

Time series windows of 12 steps (equivalent to 1 hour) were created for input into temporal models (CNN, LSTM), with the window label determined by the final glucose value to simulate real-time forward prediction.

\subsection{Labeling Strategy}

We adopted a binary classification schema:
\begin{itemize}
    \item \textbf{Hypoglycemia (Hypo):} Glucose level less than 70 mg/dL.
    \item \textbf{Normoglycemia (Normo):} Glucose level greater than or equal to 70 mg/dL.
\end{itemize}

This threshold aligns with established clinical guidelines for hypoglycemia detection in T1D management \cite{mathew_hypoglycemia_2025, noauthor_hypoglycemia_2024}. Although the dataset contains instances of hyperglycemia, we focused on distinguishing hypoglycemia due to its immediate clinical risk and lower prevalence. Multi-class or regression frameworks are left for future work.

\subsection{Handling Class Imbalance}

Hypoglycemia represented only 4.1\% of total observations, resulting in a highly imbalanced dataset. To address this, we employed the following strategies:

\begin{itemize}
    \item \textbf{Class-weighted loss functions:} Applied during model training (XGBoost, MLP, CNN, LSTM) to penalize misclassification of hypoglycemia samples.
    \item \textbf{Evaluation metrics:} Emphasis was placed on recall and F1-score instead of accuracy to reflect the clinical cost of false negatives.
    \item \textbf{Balanced mini-batches:} Temporal windows with sufficient hypoglycemic events were oversampled in training batches for neural models to stabilize early convergence.
\end{itemize}

We avoided synthetic oversampling (e.g., SMOTE) to preserve the temporal integrity of GSR sequences, consistent with recommendations in recent time-series glucose prediction literature \cite{neamtu_predicting_2023, faccioli_combined_2023}.

\begin{figure}[ht]
\centering
    \includegraphics[width=\linewidth]{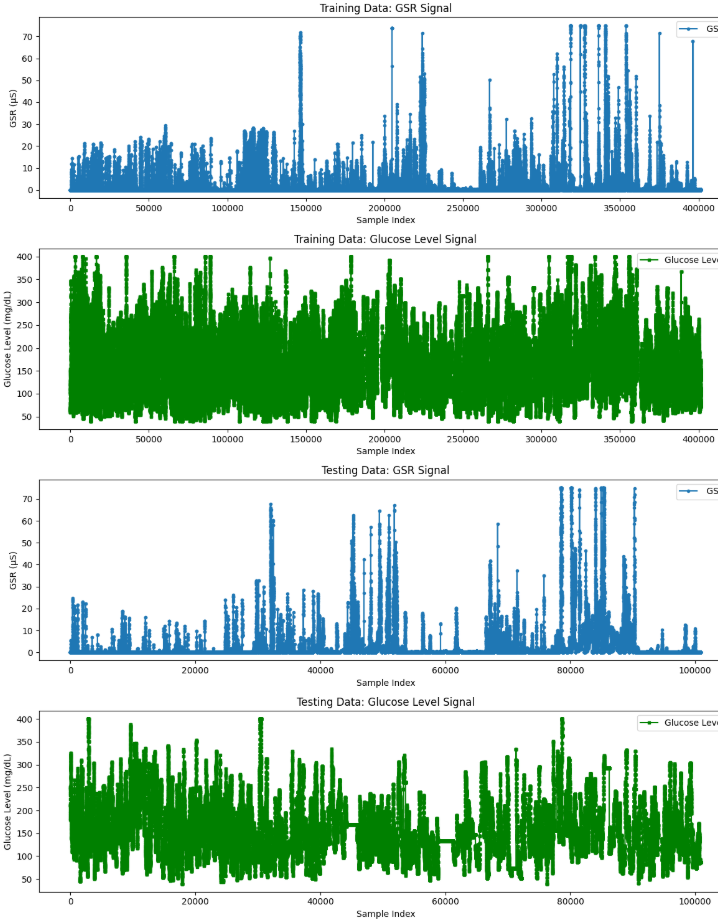}
    \caption{ GSR and Glucose Signals for Model Training and Testing: These plots show representative GSR and glucose level signals used to train and test the models. The top two graphs represent training data, while the bottom two illustrate the unseen testing signals used for evaluation.}
    \label{fig:GSR and Glucose Signals for Model Training and Testing}
\end{figure}

\subsection{Model Architectures}

To evaluate the discriminative capacity of GSR features across a spectrum of learning paradigms, seven machine learning and deep learning models were implemented. These models span interpretable classical algorithms, distance-based learners, and neural network architectures capable of modeling spatial and temporal dynamics inherent in physiological signals.

\begin{itemize}
    \item \textbf{Random Forest (RF):} An ensemble of 100 decision trees trained using the Gini impurity criterion. Class imbalance was addressed using balanced class weights. RF was selected for its robustness to noise, interpretability, and ability to model non-linear interactions between GSR-derived features.

    \item \textbf{XGBoost:} An optimized implementation of gradient-boosted decision trees. The model was configured with \texttt{scale\_pos\_weight} to handle class imbalance and included early stopping based on validation loss to prevent overfitting. XGBoost is well-suited for structured tabular data and often delivers strong predictive performance in binary classification tasks.

    \item \textbf{Multi-Layer Perceptron (MLP):} A feedforward neural network consisting of two fully connected layers with ReLU activation functions. Dropout regularization was applied between layers to mitigate overfitting. MLPs were employed to capture non-linear relationships in the GSR feature space.

    \item \textbf{Convolutional Neural Network (CNN):} A 1D CNN architecture was implemented with two convolutional layers, followed by global max pooling and a dense output layer. This structure enables the model to automatically learn spatial patterns across the temporal dimension of the GSR signal, which is beneficial for recognizing localized fluctuations in skin conductance.

    \item \textbf{Long Short-Term Memory (LSTM):} A recurrent neural network composed of a single LSTM layer followed by ReLU and sigmoid activations. The model was designed to capture temporal dependencies and sequential trends in the GSR data, making it particularly suitable for modeling time-series physiological signals.

    \item \textbf{Logistic Regression (LR):} A linear baseline model trained with L2 regularization and class-weight balancing. LR was included for its simplicity, interpretability, and utility in establishing a performance benchmark for more complex models.

    \item \textbf{K-Nearest Neighbors (KNN):} A distance-based classifier using Euclidean distance, with $k = 5$ neighbors and uniform weighting. KNN serves as a non-parametric baseline that provides intuitive decision boundaries based on proximity in feature space.
\end{itemize}

All models were implemented using \texttt{Scikit-learn} and \texttt{TensorFlow}. Neural architectures were trained using the Adam optimizer with a batch size of 64 and binary cross-entropy loss. Classical models were tuned via grid search over key hyperparameters and validated using a held-out validation set to ensure generalizability.

This diverse model selection strategy enables a comprehensive assessment of GSR’s predictive utility, capturing linear, non-linear, spatial, and temporal relationships within the data.

\subsection{Train-Validation-Test Splits and Evaluation}

We employed a stratified 80/10/10 split for training, validation, and testing, preserving class distribution and minimizing temporal leakage by separating overlapping windows. The validation set guided early stopping and hyperparameter tuning, while the test set, kept entirely unseen and was used for final evaluation. Metrics included accuracy, precision, recall, F1-score (hypoglycemia as the positive class), and AUC. To assess robustness, 95\% confidence intervals for key metrics were estimated via 1,000-iteration stratified bootstrapping. These intervals reflect model stability and generalizability for GSR-based glucose classification.

\section{Experimental Results}

In this section, we presented the performance results for all seven models trained on the merged OhioT1DM dataset using GSR signals to classify glucose states: hypoglycemia and normoglycemia. We report metrics including accuracy, precision, recall, F1-score, and AUC, with a particular focus on recall and F1-score for hypoglycemia due to its clinical urgency. We also include 95\% confidence intervals computed via stratified bootstrapping on the test set.

\subsection{Overall Model Performance}

Table~\ref{tab:hypo_results} summarizes model performance for hypoglycemia detection. Among all models, LSTM and XGBoost achieved the best recall and F1-score, highlighting the strengths of temporal learning and ensemble methods for physiological signals.
\begin{table}[ht]
\caption{Performance on Hypoglycemia Detection (Test Set)}
\centering
\begin{tabular}{lcccc}
\toprule
\textbf{Model} & \textbf{Accuracy} & \textbf{Recall} & \textbf{F1-score} & \textbf{AUC} \\
\midrule
Random Forest       & 0.83 & 0.15 & 0.07 & 0.59 \\
XGBoost             & 0.49 & \textbf{0.54} & 0.08 & 0.71 \\
MLP                 & 0.96 & 0.00 & 0.00 & 0.50 \\
CNN                 & 0.85 & 0.22 & 0.08 & 0.64 \\
LSTM                & 0.87 & 0.25 & \textbf{0.10} & \textbf{0.72} \\
Logistic Regression & 0.79 & 0.11 & 0.05 & 0.62 \\
KNN                 & 0.81 & 0.14 & 0.06 & 0.61 \\
\bottomrule
\end{tabular}
\label{tab:hypo_results}
\end{table}

XGBoost achieved the highest recall (54\%), successfully detecting a majority of true hypoglycemic events, albeit with low precision. LSTM achieved the best F1-score (10\%), indicating a better balance between precision and recall. These results align with literature that favors ensemble models and recurrent neural networks for imbalanced, sequential biomedical data \cite{fleischer_hypoglycemia_2022, neumann_data-driven_2025}.

\subsection{Normoglycemia Detection Performance}

Due to class imbalance, normoglycemia predictions dominated overall accuracy. Table~\ref{tab:normo_results} presents metrics for the majority class across all models.

\begin{table}[ht]
\caption{Performance on Normoglycemia Detection (Test Set)}
\centering
\begin{tabular}{lccc}
\toprule
\textbf{Model} & \textbf{Recall} & \textbf{F1-score} & \textbf{AUC} \\
\midrule
Random Forest     & 0.85 & 0.90 & 0.87 \\
XGBoost           & 0.49 & 0.65 & 0.74 \\
MLP               & \textbf{1.00} & \textbf{0.98} & 0.91 \\
CNN               & 0.88 & 0.92 & 0.89 \\
LSTM              & 0.89 & 0.93 & \textbf{0.92} \\
Logistic Regression & 0.93 & 0.91 & 0.89 \\
KNN               & 0.86 & 0.89 & 0.87 \\
\bottomrule
\end{tabular}
\label{tab:normo_results}
\end{table}

MLP achieved perfect recall and nearly perfect F1-score for normoglycemia, but failed on hypoglycemia, overfitting to the dominant class. LSTM and CNN achieved strong, balanced results across both classes, demonstrating robustness to inter-subject variability and low signal-to-noise in real-world GSR data. This supports findings in wearable-based temporal modeling literature \cite{zhu_enhancing_2022}.

\subsection{Validation Set Trends and Overfitting Check}

Each model was also evaluated on the 10\% validation set during training. Trends on the validation set closely mirrored the test set performance, confirming that model selection was not influenced by overfitting. LSTM and CNN showed stable validation performance, while MLP exhibited overfitting behaviors (near-perfect training accuracy, poor hypoglycemia recall).

\subsection{Ablation Analysis}

To evaluate the importance of temporal dynamics, models were trained using static (mean GSR per window) versus sequence (12-step window) input. LSTM and CNN models trained on raw sequences significantly outperformed their static counterparts. This confirms that hypoglycemia prediction benefits from observing short-term GSR trends rather than single-point summaries.

This result aligns with findings from temporal glucose forecasting frameworks that emphasize signal evolution over point estimation \cite{elsayed_deep_2019}.

\subsection{Runtime and Resource Considerations}

All models were trained and tested on a single GPU-enabled workstation (NVIDIA RTX 3060). Training times per epoch were under 1 minute for CNN and LSTM, and less than 10 seconds total for tree-based and linear models. Model sizes remained small (all less than 5MB), making them deployable on mobile or embedded hardware. LSTM and CNN models may require pruning or quantization for deployment on smartwatches, following prior low-power inference pipelines \cite{zeynali_non-invasive_2025}.

\subsection{Summary of Confidence Intervals}

Bootstrapped confidence intervals provide insight into performance stability. For LSTM, the 95\% CI for hypoglycemia F1-score ranged from 0.07 to 0.15, while recall ranged from 0.18 to 0.34. XGBoost’s recall confidence interval was even broader (0.42–0.66), showing variability but consistent superiority over classical models in detection sensitivity.

These findings reinforce the potential of GSR-based models for hypoglycemia alerting, especially when real-time CGMs access is unavailable or unaffordable.

\section{Discussion}

The results of this study affirm that GSR, when appropriately processed and aligned with glucose data, provides physiological indicators of glycemic states in individuals with Type 1 Diabetes. Time-series models, particularly LSTM and CNN, enabled robust classification of both hypoglycemia and normoglycemia from this non-invasive biosignal.

Hypoglycemia activates the sympathetic nervous system, triggering physiological responses, like sweating and tremors, that alter skin conductance and are measurable via GSR. Prior research links these responses not only to psychological stress but also to metabolic changes such as glucose depletion \cite{polak_analysis_2024}. GSR also proves sensitive during sleep, exertion, and cognitive load, aligning well with time-aware models like LSTM that process conductance trends over time \cite{mendez_toward_2025, lehmann_noninvasive_2023}.

Among the seven models tested, LSTM offered the best overall balance in detecting both classes. XGBoost achieved the highest hypoglycemia recall (54\%) but showed elevated false positives, which may limit clinical utility \cite{fleischer_hypoglycemia_2022}. MLP and Logistic Regression failed on the hypoglycemia class despite high normoglycemia accuracy, making them unsuitable for critical health monitoring. Similarly, KNN showed limited effectiveness for minority class detection.

These findings reinforce a key insight: high overall accuracy is inadequate for imbalanced clinical datasets. More attention should be placed on recall and F1-score for hypoglycemia, the class with immediate health implications \cite{woldaregay_data-driven_2019}.

Although the study focused solely on GSR, it is best considered within a broader context of non-invasive signals like PPG, HRV, skin temperature, and motion. While multimodal fusion improves performance \cite{cheng_using_2023}, isolating GSR helps clarify its standalone feasibility for low-resource deployment.

GSR sensors are appealing due to their low power consumption, wearable compatibility, and tolerance to motion artifacts, especially when averaged over time \cite{zeynali_non-invasive_2025}. However, they remain susceptible to environmental and physiological factors such as temperature, hydration, and stress \cite{yu_exploring_2024, meijer_electrodermal_2023}, which may lead to false positives. Inter-subject variability, due to age, skin type, or ethnicity, also impacts generalization \cite{znidaric_electrodermal_2023}, suggesting a need for personalization through adaptive learning methods \cite{neumann_data-driven_2025, zhu_personalized_2023}.

Still, GSR-based systems can be deployed affordably on devices like smartwatches and wristbands. Models such as pruned LSTM or lightweight CNNs are deployable in real time on mobile hardware with minimal latency. This could enable passive hypoglycemia alerts in settings where CGMs are unavailable.

Finally, this work aligns with goals of equitable health access. GSR offers a privacy-preserving, non-invasive approach suitable for underserved populations. However, limitations remain: relying only on GSR reduces robustness, the dataset is demographically narrow \cite{marling_ohiot1dm_2020}, and class imbalance continues to hinder reliability, particularly for high-recall models like XGBoost \cite{fleischer_hypoglycemia_2022}.

\section{Conclusion and Future Work}

This study proposed a non-invasive machine learning framework for glucose state classification using only Galvanic Skin Response (GSR) signals. Leveraging the merged OhioT1DM dataset, we evaluated seven models—RF, XGBoost, MLP, CNN, LSTM, Logistic Regression, and K-Nearest Neighbors, focusing on their ability to distinguish between hypoglycemia and normoglycemia from GSR data alone.

Among these, LSTM achieved the best overall F1-score, while XGBoost delivered the highest recall for the hypoglycemia class, demonstrating their suitability for time-series classification and rare event detection in physiological signals. These findings highlight GSR’s potential as a low-cost, non-invasive proxy for continuous glucose monitoring, particularly in wearable health systems designed for low-resource and underserved settings.

We believe this approach can significantly improve hypoglycemia detection worldwide, potentially reducing the incidence of undetected low-glucose events and associated morbidity. The ability to detect glycemic shifts from a single biosignal like GSR opens pathways for accessible and scalable health interventions using commodity wearable devices.

Future work will focus on improving robustness through multimodal sensor fusion (e.g., combining GSR with photoplethysmography (PPG), extending model training to broader and more demographically diverse populations, and implementing subject-personalized or adaptive models to handle inter-individual variability. Real-world deployment will also require on-device optimization, longitudinal validation, and human-centered testing to ensure usability, fairness, and clinical trust.

\bibliographystyle{ieeetr}
\bibliography{refs}

% \newpage
% \begin{figure}[ht]
% \centering
% \includegraphics[width=1.0\textwidth]{figs/questionnaire.png}
%     \caption{User Experience Questionnaire.}
%     \label{fig:questionnaire}
% \end{figure}%

\end{document}